\documentclass{article}
\usepackage[utf8]{inputenc}

\usepackage{spconf,amsmath,graphicx}
\usepackage[]{algorithm2e}
\usepackage{amsmath}
\usepackage{listings}
\usepackage{csvsimple}
\usepackage{booktabs}
\usepackage{pgfplotstable}
\DeclareMathOperator*{\argmin}{arg\,min}

\title{Low Latency Time Domain Multichannel Speech
and Music Source Separation} 
\name{Gerald Schuller}
\address{Ilmenau University of Technology,\\ 98693 Ilmenau,\\ Germany}
\begin{document}
%\ninept
%
\maketitle
\begin{abstract}
The Goal is to obtain a simple multichannel source separation with very low latency. Applications can be teleconferencing, hearing aids, augmented reality, or selective active noise cancellation. These real time applications need a very low latency, usually less than about 6 ms, and low complexity, because they usually run on small portable devices. For that we don’t need the best separation, but “useful” separation, and not just on speech, but also music and noise. Usual frequency domain approaches have higher latency and complexity.  Hence 
we introduce a novel probabilistic optimization method which we call "Random Directions", which can overcome local minima, applied to a simple time domain unmixing structure, and which is scalable for low complexity. 
Then it is compared to frequency domain approaches on separating speech and music sources, and using 3D microphone setups.
\end{abstract}
\section{Introduction}
\label{sec:intro}
The goals of this paper are to obtain a simple multichannel source separation with very low latency.
Applications for it can be teleconferencing, hearing aids, augmented reality, or selective active noise cancellation.
These real time applications need a very low algorithmic delay, usually less than about 6 ms.
Also those applications need separation not just for speech, but also for music and noise.

Previous approaches usually use the frequency domain for separation, usually using the Short Time Fourier Transform (STFT).
Their STFT typically has a block size of 4096, and hop size of 2048 samples.
The hop size alone leads to an algorithmic delay of 2047 samples.
This is 128ms at 16 kHz sampling rate, which is too much for our applications.
Examples for the frequency domain separation are FastMNMF \cite{FastMNMF} or AuxIVA \cite{AuxIVA}.

To obtain the least possible algorithmic delay, the time domain should be used for separation instead, which means to use time-domain unmixing filters, like "Trinicon" \cite{TriniconPRA}, which uses FIR unmixing filters. 
This leads to objective functions for finding the systems parameters which can be highly non-convex, for which the usual gradient based methods usually fail. 
Hence the presented approach uses a zeroth-order optimization method instead, the method of Random Directions, which is more robust against local minima. 
A difference in practice is that probabilistic methods don't have exactly the same results every time they are executed, hence the computation of the standard deviations of the results is also important.

\section{New Approach}
The presented new approach uses the time domain, with
a fractional delay IIR filters \cite{FractionalDelay} and attenuation factors. 
This models the propagation delays of the sources between the microphones, together with an attenuation factor. These fractional delays and attenuation factors are the unmixing coefficients which the optimization has to find.
%For low delay, online processing is used, which means computing the corresponding unmixing coefficients while processing the audio signal in parallel.

\subsection{Parallel Processing}
The determination of suitable unmixing filters takes some signal time.
To avoid this causing a delay, it can be computed in a parallel thread, while already processing the signal in the time domain.
This means the separation in the beginning is not at its best, but it improves quickly as more of the signal is processed.
In this way an algorithmic delay in the order of the fractional delays between the microphones is possible.

These are the two parallel threads,
\begin{itemize}
	\item unmixing in the time domain,
	\item periodic optimization of the unmixing coefficients in a parallel thread.
\end{itemize}

\subsection {Signal Accumulation}
The processing complexity for the unmixing coefficients increases with the signal length.
To reduce this complexity, the signal can be "accumulated" in a short signal block of about 0.5s length.
The assumption for it is that the solution for the coefficients is not changed by this pre-processing. It can even argued that this superposition of the signal with itself may help the optimization, since it makes the signal more broadband.
This is shown in the following code:\\
{\small
	\begin{verbatim}
	blocksize=8000; blockno=0
	for i in range(min(blocks,16)): 
	#accumulate audio blocks over ca. 3 sec:
	blockaccumulator=0.98*blockaccumulator +  
	0.02*X[blockno*blocksize+np.arange(blocksize)]
	blockno+=1
	\end{verbatim}}

\subsection{The Unmixing Function}
Let $S$ be the number of sources, $M$ be the number of microphones, $X_i(z)$ be the i-th microphone signal in the z-domain, $Y_i(z)$: the $i$-th separated source, with $a$ being the attenuation factors and $d$ the fractional delays, and with the signal vectors,
\[
{\bf X}= \left[X_1(z),\ldots, X_M(z)\right], 
{\bf Y}=\left[Y_1(z),\ldots, Y_S(z)\right] 
\]
Then the unmixing function of the presented method is,

\begin{align} 
  {\bf X} \cdot \begin{bmatrix}
      a_{1,1} z^{-d_{1,1}} & ...&a_{1,S} z^{-d_{1,S}} \\
      & \vdots &\\
      a_{M,1} z^{-d_{M,1}} & ...&a_{M,S} z^{-d_{M,S}}
      \end{bmatrix} ={\bf Y}
\end{align}
This models the delays and attenuations from a source signal between the different microphones. Observe that it does not model the room impulse response, hence the effect of the room on the sound is not canceled. 
   
\subsection{The Objective Function}
The objective function for the optimization should be a measure of statistical independence between the separated sources. Mutual information would measure it, but is complex to compute. Instead, the negative Kullback-Leibler divergence between all pairs of separated sources is used.
The shown example has only one source pair, for simplicity.
%To avoid trivial solutions, like a separated signal being very small, the preservation of power as side-condition is included using a Lagrange multiplier.
%Our objective function should measure how much mutual information the separated signals $S_i(z)$ have. 
Since a possible minimum of the objective function is also obtained by setting an output source simply to zero, or very small values, we also include a power preservation coefficient $P= |1-power_{in}/power_{out}|$.  The Kullback-Leibler Divergence $D_{KL}$ indicates how different 2 signals are, hence we want to minimize its negative value for the output sources.
A Lagrange multiplier is used to combine those two functions into one objective function $f=-D_{KL}+ \lambda \cdot P$. The Lagrange multiplier $\lambda$ was set to 0.1.

The objective function ´together with the unmixing function represents a very non-convex objective function with many local minima. Hence the zeroth-order optimization of \cite{ProbOptim} is used, which turns out to be relatively robust against local minima, and which is described next.

\section{Zeroth-Order Optimization}
Zeroth-order optimization uses no gradient, unlike the commonly used Gradient Descent algorithm, or its use in the LMS algorithm. Gradient Descend can be seen as first-order optimization, with its use of the gradient, and is robust and fast for convex objective functions, but for non-convex objective functions it will become stuck at the nearest local minimum. That is why, for the presented system, Gradient Descent does not work, and an alternative is needed. Here, the Gradient is not really helpful, and it could as well be replaced by a random vector over a given search space. This then leads to zeroth-order optimization.
Zeroth-order optimization can basically be used where Gradient Descent was used, like for online optimization. An overview can be found in \cite{ZerothOrderSPM}.
Examples are also Random Pursuit \cite{RandomPursuitETH2012}, and Gradientless Descent \cite{GLDGoogleBrain2020}.

\subsection{The Method of Random Directions}
The proposed method of Random Directions has the advantages that
it is realtively fast and has a robust convergence for our applications;
It is not increasing the objective function, hence its application does not
make things worse for difficult objective functions;
and it is still fast for higher coefficient dimensions.

In the following  pseudo code, Algorithm \ref{randdir}, shows the method of Random Directions. In the beginning it is initialized with a random starting vector $x_0$, the desired number of iterations $T$, and parallel processes $P$. 
Then there is a "scale" parameter. This defines the size of the search area, the standard deviation of the random search vector. In the beginning of the iterations this has a larger value, starting with "startingscale". In the presented experiment it was 4.0. The scale of the search vector is reduced over the duration of the iterations in a non-linear way, until it reaches "endscale", which was chosen as 0.0 in the presented experiment. 
Next there is a parallel search using $P$ randomly generated search vectors $v_p$. This is done on a random subspace of coefficients, to make it more robust for higher dimensions. Non-adaptive random subspaces are used, here of dimension 8. This means only 8 entries of $v_p$ are non-zero.

From these search vectors, the index of the best vector is obtained, and then tested if it found a new minimum of the objective function $f$. 
If so, the algorithm uses a coarse line search for the successful search vector. It was found that this speeds up convergence in more smooth areas of the objective function.
See also \cite{ProbOptim}. A simpler version for 2-channel source separation was described in \cite{StereoAudioSource,MovingSource}.

\begin{algorithm}

 %\SetKw{KwParallel}{Parallel Processing}
 \SetKwProg{KwParallel}{Parallel Processing}{:}{}
 \KwIn{$f(x)$: ${\rm I\!R}^n\rightarrow {\rm I\!R}$: Objective function to be minimized\\ 
 $x_0$: Starting point vector,\\ 
 T: Number of iterations,\\
 Number of parallel processes, \\
 startingscale: Standard deviation at the start\\ 
 endscale: Standard deviation at the end}
 Initialization:
 $x_{best}=x_0$\\
 \For{m=0,...,T}
 {$scale=endscale+(startingscale-endscale)\cdot((1.0-m/iterations)^2)$\\

 \KwParallel{}{
 {generate search vectors} $v_p$ \textrm{with std deviation}
"scale" and zero mean on \textbf{random subsets} of coefficients;\\ 
\textrm{compute} $f(x_{best}+v_p)$}
 $p_{best}= \argmin_p \{f(x_{best}+v)\mid v=v_p\}$\\
 \If{$f(x_{best}+v_{p_{best}})< f(x_{best})$} 
 {Find new $x_{best}$ with coarse line search along successful vector $v_{p_{best}}$,\\
 $k_{best} = arg min_k {f(x_{best}+2^k·v_{p_{best}})}$, $k=(-8,\ldots,8)$ (Line search)\\
 $ x_{best}= x_{best}+ 2^{k_{best}}·v_{p_{best}} $
 }
}
\KwRet{$x_{best}$}
\vspace{0.1cm}
\label{randdir}
\caption{The method of Random Directions.}
\end{algorithm}

Note that the algorithm only makes an update if the objective function became smaller. This ensures that the application of this method can only improve, but not degrade the performance, an important property for hard to optimize objective functions.
The number of iterations $T$ is chosen according to the power of the available processors. The higher the better, but it still needs to be able to run in real-time. In the presented experiment $T=1000$ was used.
For the number of parallel processes usually the available number of processors (CPU's) can be chosen. In the presented experiments it was $P=8$.

\subsubsection{Probabilistic Background}
To see how this algorithm works from the probabilistic viewpoint, let's assume $x_0$ is the current starting vector or starting point, and $v_p$ a random search vector, drawn from a symmetrical independent Gaussian probability distribution (the covariance is a diagonal matrix) with zero mean, a given standard deviation $\sigma$ (the "scale"), and with $N$ the dimension of the search subspace,
\begin{equation}
p({ v_p}) = \frac{1}{ \sigma^N \sqrt{ (2 \pi )^N}} \cdot  e^{- \frac{1}{2} (\frac{|v_p|}{\sigma})^2 }.
\label{eq:gauss}
\end{equation}
Let's further assume $v_*$ is the vector to the true absolute minimum, hence at position $ x_0+v_*$.
Let $V$ be a region around $v_*$ such that the objective function in that region is smaller than its current value (assuming it is not already in a global minimum). Hence if $v_p \in V$, then $f(x_0+v_p) < f(x_0)$.
To estimate the probability of success with a search vector, it is assumed that $p({v_p})$ is approximately constant over our relatively small region $x_0 + V$, and the volume of our N-dimensional region $V$ is $\mathrm{Vol}( V)$.  Then the probability of success, hitting this region, is approximately
\begin{equation}
p({v_p \in V}) \approx p({v_*}) \cdot \mathrm{Vol}(V).
\label{eq:prob}
\end{equation}
To maximize the probability of success with the random search vectors, $p({v_*})$, eq. (\ref{eq:gauss}), needs to be maximized, with a properly chosen standard deviation $\sigma$. This maximum is obtained with
\begin{equation}
    \sigma = |v_*|/\sqrt{N},
    \label{eq:sigma}
\end{equation} 
%(for instance from using Python sympy). 
This means a proper guess of the distance to the minimum gives a good estimate for the "scale" $\sigma$. Since it can be assumed that the distance to the minimum is iteratively reduced, the "scale" needs to be reduced accordingly. This is what the algorithm is doing with the shrinking of the scale from "startingscale" to "endscale".
Eq. (\ref{eq:gauss}) also shows that for $\sigma > 1/\sqrt{ (2 \pi )}$ the probability of success, eq.(\ref{eq:prob}), is reducing exponentially with the subspace dimension $N$. Hence at least in the beginning, for larger distances from the minimum, it is useful to reduce the dimension using subspaces.

%\section{Low Latency Online Processing}

\section{Comparison, Evaluation}
As comparison method "Trinicon" is chosen, because it is also a time domain method. The used implementation is of the "pyroomacoustics" Python module.
For a more precise comparison, the same parallel processing was applied. For both, the first ca. 8 seconds of the signal were used to obtain the optimized filter coefficients. These filter coefficients where then used to unmix the entire signal, for simplicity.

The used audio sources are pairs from speech, and also noise, and music:
Synthetic male speech from "espeak" \cite{espeak} ('espeakwav\_16.wav'), synthetic female speech ('espeakfemale\_16.wav'), pink noise from "csound" \cite{csound} ('pinkish16.wav'), tones from "csound" ('oscili\_test\_16.wav'), and music from \cite{freesound} ('fantasy-orchestra\_m16.wav'). 
The pairs where ('espeakfemale\_16.wav', 'espeakwav\_16.wav'), ('pinkish16.wav', 'espeakwav\_16.wav'), ('oscili\_test\_16.wav', 'espeakwav\_16.wav'), ('fantasy-orchestra\_m16.wav', 'espeakwav\_16.wav'). They have a sampling rate of 16 kHz and lengths between 6.3s and 11.8s. The average length of the pairs (where the shorter signal of a pair is zero padded to the length of the longer signal) is 8.4s. 
The shown processing times in Table (\ref{evaluation}) are from a computer with an Intel(R) Xeon(R) W-2123 CPU @ 3.60GHz, with 8 CPU's. As long as the processing time is below the signal length, the system is real-time capable. 
Table (\ref{evaluation}) shows that for the stereo case, Random Directions online has a mean processing time of 2.83s, with standard deviation of 0.16s. This is about 3 times faster. The longest processing time is observed for the case of the cube microphone setup, with mean 4.57s and standard deviation of 0.18s. This means that the clock frequency could be reduced until real time processing speed is reached. If a further reduction on hardware requirements is desired, the number of iterations and of parallel processes can be reduced, at the cost of a gradual reduction of separation performance.
Also observe that Trinicon has a shorter processing time.

\subsection{Microphone Setups and Simulated Room}
\begin{figure}[htb]
  \centering
  %\centerline{\includegraphics[clip, trim=2cm 1.9cm 0.1cm 1.7cm, width=0.5\textwidth]{roomcubemics.png}}
  \centerline{\includegraphics[ width=0.5\textwidth]{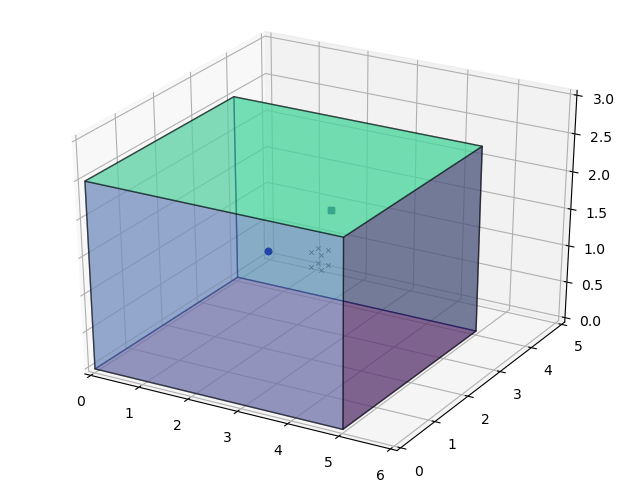}}
  \vspace{-0.2cm}
  \caption{Simulated room with cube microphone setup. The dots are the two sources, the crosses are the microphones.}
  \label{fig:multichan}
\end{figure}

3 microphone setups where testet: A stereo microphone pair, 20cm apart; a square of 4 microphones, 20cm side length; and a cube of 8 microphones, 20cm side length.
The room was simulated with the Python module "pyroomacoustics" \cite{pyroomacoustics}. The room dimensions are 5m by 4m by 2.5m, and the reverberation time was chosen as rt60=0.1s. Two source where used for the evaluation, at coordinate positions [2.5m, 1.5m, 1.50m] and [2.5m, 3.3m, 1.50m]. The microphones where centered around coordinates [3.1m, 2.1m, 1.2m], Fig. (\ref{fig:multichan}).

%\vspace{-0.2cm}

\subsection{Evaluation}
As evaluation, the Python module "mir\_eval" was used \cite{mireval}.
It computes the "Signal to Distortion Ratio" (SDR, linear distortions, like filtering), the "Signal to Interference Ratio" (SIR), 
and the "Signal to Artifacts Ratio" (SAR, the non-linear distortions).
These measures are computed and then averaged over the different source pairs, and the standard deviation is computed. 
Since the presented Random Directions algorithm is probabilistic, each setup was repeated 10 times. These sets became part of the averaging and the computation of the standard deviations. 
Here the most interesting measures are the SIR, as a measure for the separation performance, and the SAR, because it measures the non-linear distortions, at which the presented algorithm should be particularly good at. The SDR measure is less interesting here, because it measures linear distortions, like a filtering effect from the room, which the presented method is not removing and was not a goal.

%\begin{table*}[t]
%\centering
%\csvreader[
%  tabular=|r*{6}{|c}|,
%  table head= \hline & SDR & SIR & SAR & Proc.Time & micsetup & method  \\ \hline,
%  late after last line=\\\hline,
%]{meanstd.csv}{}%
%{\csvcoli & \csvcolii & \csvcoliii & \csvcoliv & \csvcolv & \csvcolvi & \csvcolvii}
%\caption{Foo}
%\end{table*}

\begin{table}[t]
\centering
\begin{filecontents*}{newtable.csv}
method, ,SIR,SAR,Time,Setup
randdironline,mean,9.37,33.02,2.83,stereo
randdironline,std.dev.,4.71,8.47,0.16,stereo
randdironline,mean,8.03,31.28,3.32,square
randdironline,std.dev.,5.48,6.6,0.15,square
randdironline,mean,4.22,30.21,4.57,cube
randdironline,std.dev.,3.44,5.83,0.18,cube
trinicononline,mean,3.38,18.84,1.12,stereo
trinicononline,std.dev.,1.02,4.58,0.02,stereo
trinicononline,mean,0.98,18.06,2.03,square
trinicononline,std.dev.,0.36,1.43,0.01,square
trinicononline,mean,1.04,19.99,3.87,cube
trinicononline,std.dev.,0.63,3.52,0.03,cube
fastmnmf,mean,20.16,14.51,1.94,stereo
fastmnmf,std.dev.,15.55,10.51,0.53,stereo
fastmnmf,mean,16.6,9.88,4.77,square
fastmnmf,std.dev.,7.99,5.34,1.49,square
fastmnmf,mean,12.27,7.92,23.55,cube
fastmnmf,std.dev.,8.87,3.79,5.84,cube
auxIVA,mean,14.72,14.14,0.22,stereo
auxIVA,std.dev.,12.3,6.1,0.06,stereo
auxIVA,mean,19.7,5.54,0.49,square
auxIVA,std.dev.,12.38,2.97,0.11,square
auxIVA,mean,16.67,2.66,1.28,cube
auxIVA,std.dev.,11.28,1.77,0.27,cube
\end{filecontents*}
\csvautobooktabular{newtable.csv}
\caption{Evaluation of the source separation for different methods and setups. "Mean" is the mean over all sources, and "std.dev." their standard deviation.}
\label{evaluation}
\end{table}

%\pgfplotstableread[col sep=comma]{meanstd.csv}{\table}
%\pgfplotstabletypeset[
%    dec sep align,      % Align at decimal point
%    fixed zerofill,     % Fill numbers with zeros
%    precision=4,        % Set number of decimals
%    display columns/0/.style={precision=1}, % Change for first column (column 0)
%    ] {\table}

%\vspace{-0.2cm}
%\begin{figure}[htb]
%  \centering
%  %\centerline{\includegraphics[clip, trim=2cm 1.9cm 0.1cm 1.7cm, width=0.5\textwidth]{roomcubemics.png}}
%  \centerline{\includegraphics[ width=0.5\textwidth]{rdstereoitem.png}}
%  \vspace{-0.2cm}
%  \caption{Box plots of SDR, SIR, SAR and processing time for Random Directions for the stereo case, 
%  for an individual item pair ('espekfemale', 'espeakwav'). The boxes represent the middle two quartiles.}
%  \label{fig:rdstereoitem}
%\end{figure}

%\vspace{-0.2cm}
\begin{figure}[htb]
  \centering
  %\centerline{\includegraphics[clip, trim=2cm 1.9cm 0.1cm 1.7cm, width=0.5\textwidth]{roomcubemics.png}}
  \centerline{\includegraphics[ width=0.5\textwidth]{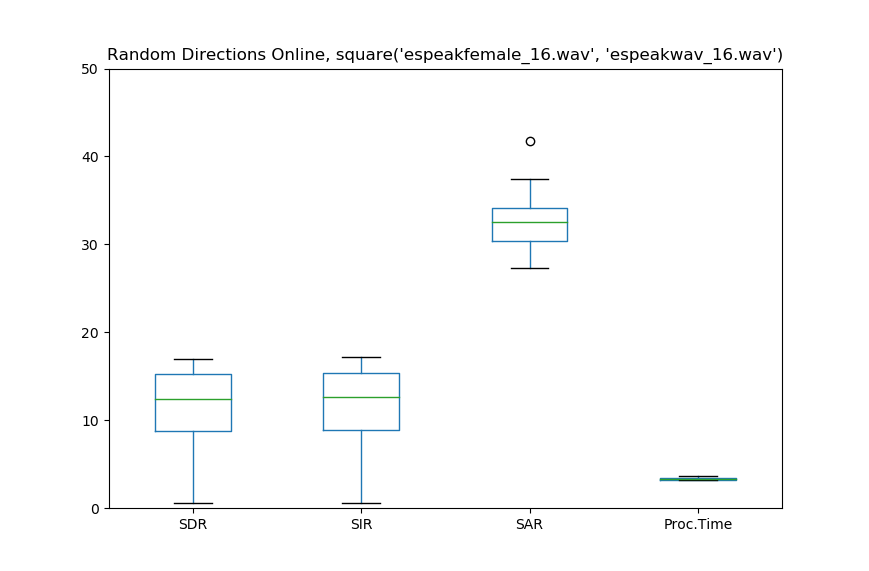}}
  \vspace{-0.2cm}
  \caption{Box plots of SDR, SIR, SAR and processing time for Random Directions for the square microphone case, 
  for an individual item pair ('espekfemale', 'espeakwav'). The boxes represent the middle two quartiles.}
  \label{fig:rdsquareitem}
\end{figure}

\subsection{Results}
Between the 2 online algorithms, the best SIR (separation) is achieved by the presented Random Directions method, with 9 dB for the stereo case. The 9 dB sound like a clear separation, with some slight audible crosstalk from the other source. 3 dB SIR, on the other hand, sounds like no audible separation.
The best overall SAR (non-lin. distortions) is achieved also with the Random Directions, with 33 dB for the stereo case. There was indeed no audible non-linear distortion.
The best overall SIR, including offline methods, is achieved by FastMNMF with 20 dB for the stereo case. In this case there is no audible crosstalk from the other source. %\cite{Githubdemo}. 
This difference can also be seen as the price for low delay separation, although both sound sufficiently separated.
Due to its stochastic nature, Random Directions also has the highest standard deviations, which means sometimes it works better than at other times. This can be also be seen in Fig. \ref{fig:rdsquareitem}, a box plot with the middle 2 quartiles for a single source pair for 10 runs of Random Directions.
It still works for more microphones (square, cubic), but declines somewhat in separation performance, but not as much as Trinicon. 
A full software demo  with listening examples in a Jupyter Colab notebook can be found at \cite{Githubsoftwaredemo}.
%\footnote{A short demo is at: https://github.com/TUIlmenauAMS/LowDelayMultichannelSourceSeparation}

\section{Conclusions}
The method of Random Directions can successfully optimize the very non-convex objective function for audio separation in the time domain for low delay online applications. It also works for non-speech signals and more microphones.
Using this zeroth-order optimization method made finding of a globally good solution possible, since the objective function is highly non-convex, which makes gradient based optimization methods fail. 
The method of Random Directions is strictly decreasing the objective function, has a line search along successful directions for increasing speed for "well behaved" objective functions, and uses a random subspace approach to make it more robust to higher dimensions, meaning more coefficients to optimize. 
The results showed that it compares favourably to Trinicon in SIR and SAR, and to frequency domain methods in terms of the SAR measure. 
%But it also has a higher standard deviation, which means that the separation results vary more, they are less consistent.

%\vfill
%\pagebreak

%\clearpage
\bibliographystyle{IEEEbib}
\bibliography{literature,optimrefs}

\end{document}